\documentclass[twocolumn,showpacs,preprintnumbers,amsmath,amssymb,superscriptaddress,nofootinbib]{revtex4-1}

\usepackage{graphicx}
\usepackage{amsmath}
\usepackage{epsfig}
\usepackage{bm}
\usepackage[usenames]{color}

\newcommand{\noun}[1]{\textsc{#1}}

\usepackage{color}

\begin{document}

\title{ Tunable Bound States in Continuum by Optical Frequency }

\author{Yingyue Boretz}
\affiliation{Center for Studies in Statistical Mechanics and Complex
Systems, The University of Texas at Austin, Austin, TX 78712 USA} 
\author{Gonzalo Ordonez}
\affiliation{Physics and Astronomy Department, Butler
University, 4600 Sunset Ave., Indianapolis, IN 46208 USA} 
\author{Satoshi Tanaka}
\affiliation{Department of Physical Science, Osaka Prefecture 
University, Gakuen-cho 1-1, Sakai 599-8531, Japan}
\author{Tomio Petrosky}
\affiliation{Center for Studies in Statistical Mechanics and Complex
Systems, The University of Texas at Austin, Austin, TX 78712 USA}

\date{\today}

\begin{abstract}
 We demonstrate  the existence of tunable bound-states in continuum (BIC)
in a 1-dimensional quantum wire with two impurities induced by an intense
monochromatic radiation field. We found that there is a new type of BIC 
due to the Fano interference between two optical transition channels, in addition
to  the ordinary BIC due to  a geometrical interference between electron wave functions emitted by impurities.
In both cases the BIC can be achieved by tuning
the frequency of the radiation field.

\end{abstract}

\maketitle
\vfill

\section{{\normalsize Introduction}}

The phenomenon of bound-states in continuum (BIC) was first discovered by Wigner and von Neumann
\cite{Von -W}. Subsequent studies can be found in a number of papers
(e.g., \cite{Sudarshan}-\cite{Sadreev}). An experimental report showed evidence
of BIC in super-lattice structures of quantum wells with a single
impurity site \cite{F.Cappasso} or in single defect tube \cite{P. S. Deo}.
Examples of BIC in a double-cavity, 2-dimensional (2D)  electron waveguide
was reported in \cite{Ordonez-Na}-\cite{Linda}.

In general if a discrete state embeds inside the continuum, the state will become
unstable due to the resonance effect. If the transition channels are
more than one, the resonance line-shape becomes asymmetric due to
the quantum interference between those decay channels. The phenomenon
often refereed as the Fano interference \cite{Fano,Fano-1-1}.
There are many studies that have  followed Fano's work. However,
the phenomena of the BIC and the Fano interference have been often
studied as individual effects. In this paper we discuss the relation between  
BIC and Fano interference. 

As an example, we consider here a tight-binding model with two
\emph{intra-atoms }attached to a semiconductor nanowire under a constant
irradiation of an intense monochromatic radiation field. An electron
of an \emph{intra-atom} is excited by the radiation field from a lower
energy state to an intermediate energy state to states with a continuous range of energies; alternatively,  the electron from the lower energy state can jump directly to the continuous states. We label these two optical transition paths as $T_{1\cdot}$ and $T_{2\cdot}$,
respectively (see FIG.1). Since there are two  optical transition channels,  Fano interference appears in this model 

\begin{figure}
\includegraphics[width=8cm]{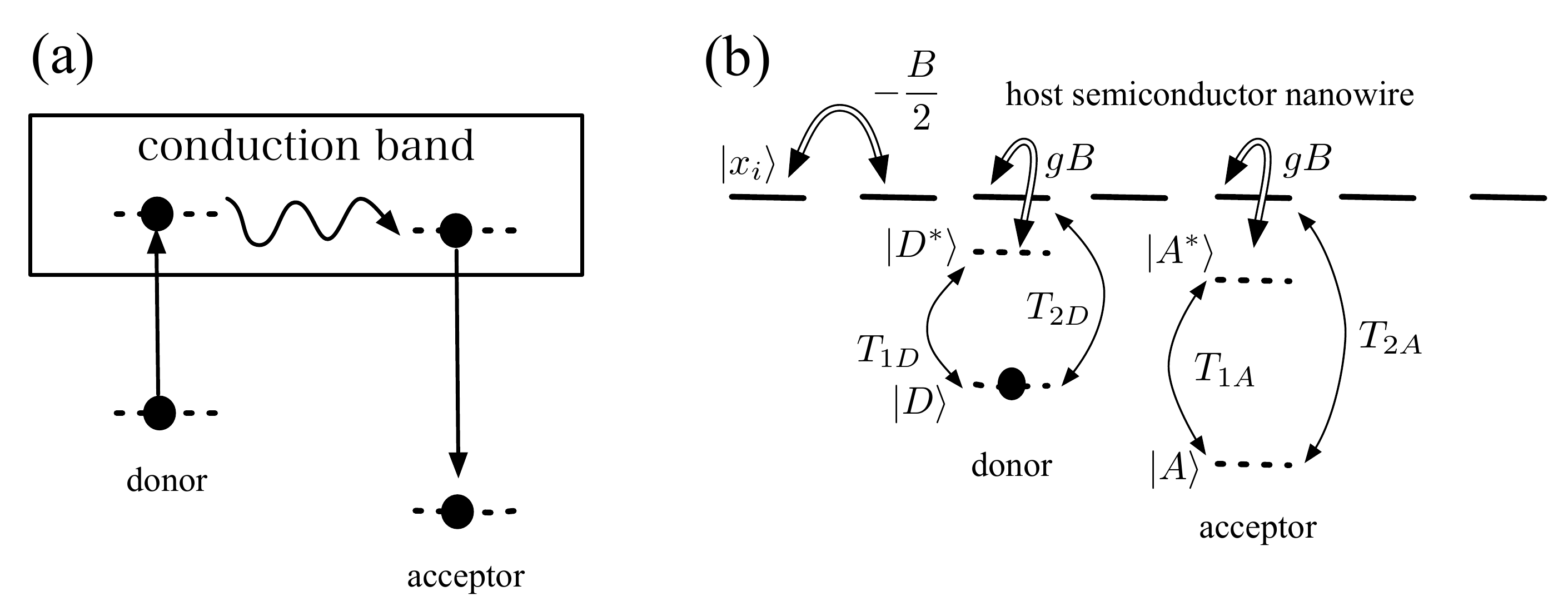}
\caption{ (a) Energy levels of donor and acceptor atoms. (b) Transitions among energy levels}
\end{figure}

The main results that we present in this paper are  twofold: The first
one is a new type of BIC in this system. 
In this BIC, the energy of the bound state
depends on the coupling constant $g$ of the interaction between
the discrete state and the continuum. This is not the case of ordinary
BIC that has been discussed before (see, e.g., Refs. \cite{Tanaka,Longhi}).
The ordinary BIC can be found for special values of energy of the
discrete state that are imbedded in the continuum, where the energy
shift of the discrete state due to the interaction vanishes. This
value of the energy may found by requiring that the so-called \textquotedblleft{}self-energy
part\textquotedblright{}of the discrete state vanishes. Hence, this
type of BIC has the same energy as the unperturbed energy without
the interaction. We found this type of BIC in our model. In addition,
however, we found the new type of BIC as mentioned above. Since this
new type of BIC depends on the interaction, we call this a \textquotedblleft{}dynamic
BIC,\textquotedblright{} while we call the ordinary type of BIC a
\textquotedblleft{}static BIC.\textquotedblright{} As discussed in
\cite{Tanaka}, the static BIC is due to a geometrical interference
in the wire between electron wave functions emitted by impurities.
\footnote{In other systems such as the system of Ref. \cite{G.Ordonez}, the energy of the BIC due to geometrical interference does depend on the interaction; however in the present system this energy is independent of the interaction. Due to this feature it is much easier to distinguish the two types of BIC in the present system.}
In contrast, the dynamic BIC appears because of the multichannels
of the transition, $T_{1}{\cdot}$ and $T_{2}{\cdot}$, as we shall show. Hence,
the dynamic BIC is the result of Fano interference.

The second main result is that because the freedom to chose the frequency
of the radiation field,  both BIC (the dynamic BIC and static BIC)
may exist for wide value of the spectrum of the discrete state. This
is not the case of the BIC that has been discussed in \cite{Tanaka}
for the system without the radiation field. Indeed, in the absence
of the radiation field, we have shown in \cite{Tanaka} that the BIC
may exist only for a special value of the discrete energy. In contrast, here
one can tune the frequency of the radiation field in order to achieve
the BIC for arbitrary value of the energy of the discrete state. This
tune-ability makes the BIC phenomena much more feasible to observe experimentally.

This paper is organized as follows. In section 2, we introduce the
model. Then, we decompose the Hamiltonian of the system into the symmetric
part and anti-symmetric part so that we can analyze our problem
in much simpler form. In section 3, we construct the complex eigenvalue
of resonance states to analyze the instability of the discrete states
inside the continuum. Then we find the BIC by requiring
that the imaginary part of the complex eigenvalue of the Hamiltonian
vanishes at the BIC. In section 4, we present several cases of the
dynamic BIC and the static BIC by plotting the imaginary part of the
eigenvalue as a function of the frequency of the radiation field.
In section 5, we summarize  our results.

\section{ Model }

We shall consider a semiconductor nanowire with donor-acceptor impurities,
e.g. 3D transition metal impurities  \cite{Pradhan11,Chin09},
where the multiplet discrete levels of transition metals appear in
the semiconductor band gap \cite{Watanabe87}. An electron of a donor
is excited by an optical transition and is transferred to the acceptor
through a semiconductor conduction band, and the electron is dexcited
by an \emph{intra-atomic} transition to emit a photon (see FIG. 1).

We show the model system of the present work in FIG. 1. The system
consists of a semiconductor nanowire with donor and acceptor impurities
located at $x_{D}$ and $x_{A}$, respectively. The semiconductor
nanowire is described by a 1D tight-binding model
with a nearest neighbor interaction $-B/2$ yielding a 1D conduction
band with bandwidth $B$ with a lattice constant of $d$. We consider
the lower and higher energy states of the donor (acceptor) impurity
represented by $|D\rangle$ ($|A\rangle$) and $|D^{*}\rangle$ ($|A^{*}\rangle$),
respectively. In this paper we use a conventional
notation ``{*}'' for excited states used in Atomic Molecular and
Optical physics. We consider the charge transfer between the higher energy
state to the nanowire at the impurity sites of $x_{D}$ and $x_{A}$
with a coupling $gB$, where $g$ is a dimensionless coupling constant.
The electronic Hamiltonian is then represented by 

\begin{eqnarray}
H_{el} & = & E_{D}|D\rangle\langle D|+E_{D^{*}}|D^{*}\rangle\langle D^{*}|\nonumber \\
 &  & +E_{A}|A\rangle\langle A|+E_{A^{*}}|A^{*}\langle A^{*}|\nonumber \\
 &  & +E_{0}\sum_{i=-N/2}^{N/2}|x_{i}\rangle\langle x_{i}|-\frac{B}{2}\sum_{<i,i'>}|x_{i}\rangle\langle x_{i'}|\nonumber \\
 &  & +gB\left(|x_{D}\rangle\langle D|+|D^{*}\rangle\langle x_{D}|\right)\nonumber \\
 &  & +gB\left(|x_{A}\rangle\langle A^{*}|+|A^{*}\rangle\langle x_{A}|\right)\;,
\end{eqnarray}
where $E_{D}$ ($E_{A}$) and $E_{D^{*}}$ ($E_{A^{*}}$)
are the energies of $|D\rangle$ ($|A\rangle$) and $|D^{*}\rangle$
($|A^{*}\rangle$), respectively. The symbol $<i,i^{\prime}>$ represents the sum over
nearest neighbors, where the sum runs from $-N$ to $N$. 

The 1D tight-binding Hamiltonian is diagonalized by the wave-number
representation defined by 
\begin{equation}
|k\rangle=\frac{1}{\sqrt{L}}\sum_{i=-N/2}^{N/2}e^{ikx_{i}}|x_{i}\rangle\;,
\end{equation}
Where under the periodic boundary condition the wave number takes
the values of 
\begin{equation}
k_{j}=\frac{2\pi j}{Nd}\;,\;\left(j=\text{interger , }-\frac{N}{2}\leq j<\frac{N}{2}\right)\,\label{eq:bc}
\end{equation}
with the length of the nanowire $L\equiv Nd$. We consider the case
$N\gg1$, and approximate it by taking the limit $N\rightarrow\infty$. In this limit we have 
\begin{equation}
\frac{2\pi}{L}\sum_{j=-N/2}^{N/2}\to\int_{-\pi/d}^{\pi/d}\, dk,\;\;\;\;\;\frac{2\pi}{L}\delta_{j,j'}^{Kr}\to\delta(k-k')\;,\label{eq: krtodi}
\end{equation}
 where the ``$\delta^{Kr}$'' stands for Kronecker delta. We will take this limit in section 3.

In terms of the wave number representation, $H_{el}$ reads 
\begin{eqnarray}
H_{el} & = & E_{D}|D\rangle\langle D|+E_{D^{*}}|D^{*}\rangle\langle D^{*}|\nonumber \\
 & + & E_{A}|A\rangle\langle A|+E_{A^{*}}|A^{*}\rangle\langle A^{*}|\nonumber \\
 & + & \sum_{k=-\pi/d}^{\pi/d}E_{k}|k\rangle\langle k|\nonumber \\
 & + & \frac{gB}{\sqrt{L}}\sum_{k=-\pi/d}^{\pi/d}\left(e^{-ikx_{D}}|k\rangle\langle D^{*}|+e^{ikx_{D}}|D^{*}\rangle\langle k|\right)\nonumber \\
 & + & \frac{gB}{\sqrt{L}}\sum_{k=-\pi/d}^{\pi/d}\left(e^{-ikx_{A}}|k\rangle\langle A^{*}|+e^{ikx_{A}}|A^{*}\rangle\langle k|\right)\;,\label{eq: hel-5}
\end{eqnarray} where the dispersion relation of an electron in the continuum is
given by 
\begin{equation}
E_{k}= E_{0}-2B\cos(kd)\;.\label{Ek}
\end{equation}
As convention\emph{ }we will use the summation notation over wave
vector $k$. In Eq.(\ref{eq: hel-5}) and hereafter. In this paper
we will set the origin of energy at $E_{0},$ i.e., $E_{0}=0,$ then
we have  $E_{k}=-2B\cos(kd).$

We also consider a monochromatic radiation field with a frequency
$\Omega$ which is close to the transition energies of $E_{D^{*}}-E_{D}$
or $E_{A^{*}}-E_{A}$. The radiation field is described by 
\begin{equation}
H_{R}=\hbar\Omega b^{\dagger}b\;,
\end{equation}
where $b$ ($b^{\dagger}$) is an annihilation (creation) operator
for the radiation field.

As for the interaction of the electron with the radiation field, we
consider two optical transition paths from the impurity lower
levels. One is the \emph{intra-atomic }transition in which an electron
is excited from the lower impurity level to the upper impurity level.
The other is the\emph{ inter-atomic} transition in which an electron
at the lower impurity level is directly excited into the host semiconductor
nanowire at the impurity site. Then the interaction Hamiltonian is
described under the dipole approximation \cite{F.H.Stillinger} as
\begin{eqnarray}
H_{V} & = & T_{1D}\left(|D^{*}\rangle\langle D|b+|D\rangle\langle D^{*}|b^{\dagger}\right)\nonumber \\
 & + & T_{1A}\left(|A^{*}\rangle\langle A|b+|A\rangle\langle A^{*}|b^{\dagger}\right)\nonumber \\
 & + & T_{2D}\left(|x_{D}\rangle\langle D|b+|D\rangle\langle x_{D}|b^{\dagger}\right)\nonumber \\
 & + & T_{2A}\left(|x_{A}\rangle\langle A|b+|A\rangle\langle x_{A}|b^{\dagger}\right)\;,\label{eq:HV}
\end{eqnarray}
where $T_{1\cdot}$ and $T_{2\cdot}$ represent the transition strengths for
the two optical transitions. Since the monochromatic radiation $\hbar\Omega$
is near resonant to the transition from the lower level to the upper
level or semiconductor conduction band, we have used rotating wave
approximation (RWA) in Eq. (\ref{eq:HV}) where we have neglected 
further excitation from the conduction electron to higher excited
states.

Even though the interactions of the electron with the radiation field,
$T_{1\cdot}$ and $T_{2\cdot}$, are small, when the the radiation field intensity
is large with a large value of $n$, we have to incorporate the radiation
field  non-perturbatively in terms of the dressed state concept.
We then consider the composite vector space of the electronic states
and the radiation field \cite{CohenTannouji}. Let us denote the number
state $|n\rangle$ ($n=0,1,2,\dots$) as an eigenstate of the radiation
filed. Then the composite vector basis is comprised of $|\alpha,n\rangle$,
where $\alpha$ denotes the electronic states: $\alpha=D,A,D^{*},A^{*}$, and $k$.
In terms of these basis, total Hamiltonian is described by 
\begin{widetext}
\begin{eqnarray}
H & =&  H_{el}+H_{R}+H_{V}\nonumber \\
 &  &= \sum_{n=0}^{\infty}\sum_{\alpha=D,A,D^{*},A^{*},k}(E_{\alpha}+\hbar\Omega n)|\alpha,n\rangle\langle\alpha,n|\nonumber \\
 &  +& \frac{gB}{\sqrt{L}}\sum_{n=0}^{\infty}\sum_{k=-\pi/d}^{\pi/d}\Big(e^{-ikx_{D}}|k,n\rangle\langle D^{*},n|+e^{ikx_{D}}|D^{*},n\rangle\langle k,n| + e^{-ikx_{A}}|k,n\rangle\langle A^{*},n|+e^{ikx_{A}}|A^{*},n\rangle\langle k,n|\Big)\nonumber \\
 &  +& \sum_{n=1}^{\infty}\sqrt{n}[T_{1D}(|D^{*},n-1\rangle\langle D,n|+|D,n\rangle\langle D^{*},n-1|)+ T_{1A}(|A^{*},n-1\rangle A,n|+|A,n\rangle\langle A,n-1|)] \nonumber\\
 & +& \sum_{n=1}^{\infty}\frac{\sqrt{n}}{\sqrt{L}}\sum_{k=-\pi/d}^{\pi/d}[T_{2D}(e^{-ikx_{D}}|k,n-1\rangle\langle D,n|+e^{ikx_{D}}|D,n\rangle\langle k,n-1|) \nonumber\\
 && + T_{2A}(e^{-ikx_{A}}|k,n-1\rangle\langle A,n|+e^{ikx_{A}}|A,n\rangle\langle k,n-1|)]   \;.
\end{eqnarray}
This can be also written as 
\begin{eqnarray}
H & = & \sum_{n=0}^{\infty}\Big\{\sum_{\alpha=D,A}\big(E_{\alpha}+\hbar\Omega(n+1)\big)|\alpha,n+1\rangle\langle\alpha,n+1|
  +  \sum_{\alpha=D^{*},A^{*},k}\big(E_{\alpha}+\hbar\Omega n\big)|\alpha,n\rangle\langle\alpha,n|\nonumber \\
 & + & \frac{gB}{\sqrt{L}}\sum_{k=-\pi/d}^{\pi/d}\Big(e^{-ikx_{D}}|k,n\rangle\langle D^{*},n|+e^{ikx_{D}}|D^{*},n\rangle\langle k,n|  +  e^{-ikx_{A}}|k,n\rangle\langle A^{*},n|+e^{ikx_{A}}|A^{*},n\rangle\langle k,n|\Big)\nonumber \\
 & + & \sqrt{n+1}[T_{1D}(|D^{*},n\rangle\langle D,n+1|+|D,n+1\rangle\langle D^{*},n|) +  T_{1A}(|A^{*},n\rangle\langle A,n+1|+|A,n+1\rangle\langle A,n|)]\nonumber \\
 & + & \frac{\sqrt{n+1}}{\sqrt{L}}\sum_{k=-\pi/d}^{\pi/d}[T_{2D}(e^{-ikx_{D}}|k,n\rangle\langle D,n+1|+e^{ikx_{D}}|D,n+1\rangle\langle k,n|)  \nonumber\\
 &&+  T_{2A}(e^{-ikx_{A}}|k,n\rangle\langle A,n+1|+e^{ikx_{A}}|A,n+1\rangle\langle k,n|)]\Big\}\nonumber \\
 & \equiv & \sum_{n=0}^{\infty}H_{n}  \;.
\end{eqnarray}
\end{widetext}
Note that the total vector subspace is classified into independent
manifolds according to the photon number $n$ \cite{CohenTannouji}.

In the present work, we solve the complex eigenvalue problem of $H$.
For simplicity, we shall consider a symmetric situation where 
\begin{eqnarray}
&&x_{D}=-x_{A}\;,\; E_{l}\equiv E_{D}=E_{A}\;,\; E_{u}\equiv E_{D^{*}}=E_{A^{*}}\;\;,\nonumber\\
&& T_{i}\equiv T_{iA}=T_{iD}\;,
\end{eqnarray}
where $l$ stands for the lower level, and $u$ stands for the upper
level. In this case, because of the inversion symmetry of the system, we
can further decompose the vector space according to the parity. We
denote the following symmetrized basis as (for symmetric basis) 
\begin{eqnarray}
|S_{l},n+1\rangle & \equiv & \frac{1}{\sqrt{2}}(|D,n+1\rangle+|A,n+1\rangle),\\
|S_{u},n\rangle & \equiv & \frac{1}{\sqrt{2}}(|D^{*},n\rangle+|A^{*},n\rangle),\\
|S_{k},n\rangle & \equiv & \frac{1}{\sqrt{2}}(|k,n\rangle+|-k,n\rangle),
\end{eqnarray}
 and (for anti-symmetric basis) 
\begin{eqnarray}
|P_{l},n+1\rangle & \equiv & \frac{1}{\sqrt{2}}(|D,n+1\rangle-|A,n+1\rangle),\\
|P_{u},n\rangle & \equiv & \frac{1}{\sqrt{2}}(|D^{*},n\rangle-|A^{*},n\rangle),\\
|P_{k},n\rangle & \equiv & \frac{1}{\sqrt{2}}(|k,n\rangle-|-k,n\rangle).
\end{eqnarray}
With these basis, $H_{n}$ is divided as 
\begin{equation}
H_{n}=H_{n}^{p}+H_{n}^{s}\;,
\end{equation}
where 
\begin{eqnarray}
H_{n}^{s} & = & \big(E_{l}+\hbar\Omega(n+1)\big)|S_{l},n+1\rangle\langle S_{l},n+1|\nonumber \\
 &  & +(E_{u}+n\Omega)|S_{u},n\rangle\langle S_{u},n|+\sum_{k=-\pi/d}^{\pi/d}E_{k}|S_{k},n\rangle\langle S_{k},n|\nonumber \\
 & + & \frac{gB}{\sqrt{L}}\sum_{k=0}^{\pi/d}2\cos(kx_{D})\Big(|S_{k},n\rangle\langle S_{u},n|+|S_{u},n\rangle\langle S_{k},n|\Big)\nonumber \\
 & + & T_{1}\sqrt{n+1}\Big(|S_{u},n\rangle\langle S_{l},n+1|+|S_{l},n+1\rangle\langle S_{u},n|\Big)\nonumber \\
 & + & T_{2}\frac{\sqrt{n+1}}{\sqrt{L}}\sum_{k=0}^{\pi/d}2\cos(kx_{D})\Big(|S_{k},n\rangle\langle S_{l},n+1|\nonumber \\
 &  & +|S_{l},n+1\rangle\langle S_{k},n|\Big)\;,\label{eq:hs}
\end{eqnarray}
and 
\begin{eqnarray}
H_{n}^{p} & = & \big(E_{l}+\hbar\Omega(n+1\big))|P_{l},n+1\rangle\langle P_{l},n+1|\nonumber \\
 &  & +(E_{u}+n\Omega)|P_{u},n\rangle\langle P_{u},n|+\sum_{k=-\pi/d}^{\pi/d}E_{k}|P_{k},n\rangle\langle P_{k},n|\nonumber \\
 & + & \frac{gB}{\sqrt{L}}\sum_{k=0}^{\pi/d}2i\sin(kx_{D})\Big(|P_{k},n\rangle\langle P_{u},n|+|P_{u},n\rangle\langle P_{k},n|\Big)\nonumber \\
 & + & T_{1}\sqrt{n+1}\Big(|P_{u},n\rangle\langle P_{l},n+1| - |P_{l},n+1\rangle\langle P_{u},n|\Big)\nonumber \\
 & + & T_{2}\frac{\sqrt{n+1}}{\sqrt{L}}\sum_{k=0}^{\pi/d}2i\sin(kx_{D})\Big(|P_{k},n\rangle\,\langle P_{l},n+1|\nonumber \\
 &  &  - |P_{l},n+1\rangle\langle P_{k},n|\Big)\;.\label{eq:hp}
\end{eqnarray}

Hereafter we use the units $d=1$ and $\hbar=1.$ By taking the limit
$L\equiv Na\rightarrow\infty$, summation over the wave number
$k$ turns into the integration as in Eq.(\ref{eq:bc}).

\section{{\normalsize Optical Dressed Bound State in Continuum}}

As we pointed out previously, our main focus is to study the decay
process under influence of a constant irradiation of an intense monochromatic
optical field. For this purpose, we solve the complex eigenvalue problem
of the Hamiltonian. The solutions corresponding to unstable state
are found on the second Riemann sheet of the complex energy plane.
The imaginary part gives decay rate of the unstable state. 

We shall solve the complex eigenvalue problem of the Hamiltonian:
\begin{equation}
H_{n}\psi_{E}=E\psi_{E}\label{eq:Hpsi} .
\end{equation}

We  start with the anti-symmetric sector, i.e., $p$-sector
in Eq.(\ref{eq:hp}). We denote the components of the eigenstates in the $p$-sector
 as 
\begin{equation}
\left(\begin{array}{c}
\tilde{D}\\
\tilde{D}^{*}\\
\tilde{x}_{k}
\end{array}\right)\equiv\left(\begin{array}{c}
\langle P_{l},n+1|\psi_{E}\rangle\\
\langle P_{u},n|\psi_{E}\rangle\\
\langle P_{k},n|\psi_{E}\rangle
\end{array}\right) ,
\end{equation}

From Eq. (\ref{eq:Hpsi}) we obtain the following system of equations (for
$d=1$) 
\begin{widetext}
\begin{eqnarray}
(E_{l}+(n+1)\Omega)\tilde{D}+\sqrt{n+1}T_{1}\tilde{D}^{*}+\frac{\sqrt{n+1}T_{2}}{\pi}\int_{-\pi}^{\pi}dk\, i\text{sin}(x_{D}k)\tilde{x}_{k} & = & E\tilde{D} \;,\nonumber \\
\sqrt{n+1}T_{1}\tilde{D}+(E_{u}+n\Omega)\tilde{D}^{*}+\frac{g}{\pi}\int_{-\pi}^{\pi}dk\, Bi\text{sin}(x_{D}k)\tilde{x}_{k} & = & E\tilde{D}^{*}\;,\nonumber \\
\text{\ensuremath{-\frac{\sqrt{n+1}T_{2}}{\pi}i}sin}(x_{D}k^{\prime})\tilde{D}-\text{\ensuremath{\frac{gB}{\pi}i}sin}(x_{D}k^{\prime})\tilde{D}^{*}+\frac{1}{\pi}\int_{-\pi}^{\pi}dk\, B\text{cos}(x_{D}k)\delta(k-k^{\prime})\tilde{x}_{k} & = & E\tilde{x}_{k}\;.\nonumber \\
\end{eqnarray}
\end{widetext}

From the above relations, we obtain the eigenvalue equation for $p$-sector.
With similar calculations, we can also obtain the eigenvalue equations
for the symmetric sector, $s$-sector in Eq.(\ref{eq:hs}). We summarize
 both $p$- and $s$-sectors into one form as the following
eigenvalue equations whose solutions give the resonant-state pole
of the resolvent operator $[z-H_{n}]^{-1}$ at $z=E$ in the second Riemann
sheet,
\begin{widetext}
\begin{eqnarray}
 && (z-((n+1)\Omega+E_{l}))(z-(E_{u}+n\Omega))-(n+1)T_{1}^{2}\nonumber \\
 && -\Xi^{p,s}(z)g^{2}[(z-(\Omega(n+1)+E_{l}))B+2(n+1)\frac{T_{1}T_{2}}{g}+(n+1)\frac{T_{2}^{2}}{g^{2}B}(z-(E_{u}+n\Omega))] =0\nonumber \\
\label{eq:disp}
\end{eqnarray}
\end{widetext}
where $\Xi^{p,s}(z)$ are the self-energies of the Hamiltonian that
without the lower energy level and external radiation field \cite{Tanaka}
\begin{eqnarray}
&&\Xi^{p,s}(z)  \equiv\frac{1}{\pi}  \int_{-\pi}^{\pi}dk\frac{B(1\pm\text{cos\ensuremath{(2kx_{D}))}}}{(z-B\mbox{cos\ensuremath{k)}}}\nonumber \\
 && =  \frac{1}{i\sqrt{1-z^{2}/B^{2}}}\left[1\pm\left(-\frac{z}{B}+\sqrt{1-z^{2}/B^{2}}\right)^{2x_{D}}\right]\label{eq: self1}
\end{eqnarray}
where the plus and minus is for the $s$- and $p$-sectors, respectively.
Putting 
\begin{equation}
z=-B\text{cos\ensuremath{\theta}}\label{eq:dis-e}
\end{equation}
we have 
\begin{equation}
\Xi^{p,s}(z)=\frac{1}{i\text{sin\ensuremath{\theta}}}(1\pm e^{i2x_{D}\theta})
\end{equation}

The BIC corresponds to real solution of the eigenvalue Eq.(\ref{eq:disp}). Note that if the last term of the equation vanishes,
we obtain 
\begin{eqnarray}
z&=&\frac{1}{2}\Big\{\left((2n+1)\Omega+E_{l}+E_{u}\right)\nonumber\\
&\pm &\sqrt{(\Omega+E_{l}-E_{u})^{2}+4(n+1)T_{1}^{2}}\Big\}\label{eq:z}
\end{eqnarray}
which are real solutions. 

One can show that these are the only real solutions of Eq.(\ref{eq:disp})
as follows: Let us denote the real eigenvalue as 
\begin{equation}
z=z_{0}
\end{equation}
Substituting it into Eq.(\ref{eq:disp}), we have 
\begin{widetext}
\begin{eqnarray}
 &  & (z_{0}-((n+1)\Omega+E_{l}))(z_{0}-(n\Omega+E_{u}))-(n+1)T_{1}^{2}\nonumber  \\
 &  &= \Xi^{p,s}(z_{0})[(z_{0}-((n+1)\Omega+E_{l}))Bg^{2}+(n+1)2gT_{1}T_{2}+(n+1)\frac{T_{2}^{2}}{B}(z_{0}-(E_{u}+n\Omega))]\nonumber \\
\label{eq:proof of two solutions}
\end{eqnarray}
\end{widetext}
Note the left-hand side itself and the factor in front of $\Xi^{p,s}(z_{0})$
are both real, because all parameters are real. Hence $\Xi^{p,s}(z_{0})$
must be real, or else, the factor in front must vanish. 

By the definition of BIC we have 
\begin{equation}
|\frac{z_{0}}{B}|\leq1
\end{equation}
 As a result $\theta$ in Eq.(\ref{eq:dis-e}) is real for $z=z_{0}$.
Therefore, $\Xi^{p,s}(z_{0})$ is a complex number with a non-vanishing
imaginary part except for 
\begin{equation}
\Xi^{p,s}(z)=\frac{1}{i\text{sin\ensuremath{\theta}}}(1\pm e^{i2x_{D}\theta})=0\label{eq:sef}
\end{equation}
 Eq. (\ref{eq:sef}) leads to one possible set of the
BIC that satisfies 
\begin{equation}
1\pm e^{i2x_{D}\theta}=0\label{eq:dx=00003D0}
\end{equation}
Then, this leads to Eq.(\ref{eq:z}).

On the other hand, if Eq.(\ref{eq:sef}) is not satisfied, then,
$\Xi^{p,s}(z)$ is a complex number as mentioned above. Hence, to
be consistent which the fact that the left-hand side of Eq.(\ref{eq:proof of two solutions})
must be real, we shall have 
\begin{eqnarray}
&&(z_{0}-((n+1)\Omega+E_{l}))Bg^{2}+(n+1)2gT_{1}T_{2}\nonumber\\
&&+(n+1)\frac{T_{2}^{2}}{B}(z_{0}-(E_{u}+n\Omega))=0\label{eq: dz}
\end{eqnarray}

Hence, once again we obtain Eq.(\ref{eq:z}). This proves that $z$ in Eq.(\ref{eq:z})
are only the real solutions of Eq.(\ref{eq:disp}).

Let us first consider the case of Eq.(\ref{eq:sef}). We notice that
the self-energy for the $s$- and $p$-sectors periodically vanish
when 
\begin{equation}
\theta=\frac{m\pi}{2x_{D}}\,,\,\,\begin{cases}
\text{even integer \ensuremath{m} for \ensuremath{p}-sector}\\
\text{odd interger \ensuremath{m} for \ensuremath{s}-sector}
\end{cases}
\end{equation}
and then the real solution of the eigenvalue equation, i.e. BIC, is
given
\begin{equation}
z_{0}=-B\cos\left(\frac{m\pi}{2x_{D}}\right)\label{eq:z_0&m}
\end{equation}
Note that the energies of the BIC are the same as obtained in \cite{Tanaka},
where $z_{0}$ does not depend on $g$. This is a typical feature
of the ordinary BIC in this system, hence the static BIC mentioned in the introduction
comes from a geometrical interference of the two electron wavefunctions emitted  from $|D^{*}\rangle$ 
and $|A^{*}\rangle$ states.

Substituting Eq. (\ref{eq:z_0&m}) into Eq.(\ref{eq:proof of two solutions}) with the right-hand-side equal $0$, we obtain
an equation for the frequency $\Omega$ of the photon which can achieve static BIC in this
system 
\begin{eqnarray}\label{eq:Om}
&&\left[-\cos\left(\frac{m \pi}{2 x_D}\right) -((n+1)\Omega+E_{l})\right] \nonumber\\
&\times& \left[-\cos\left(\frac{m \pi}{2 x_D}\right)
- (n\Omega+E_{u})\right]-(n+1)T_{1}^{2} =0\nonumber\\
\end{eqnarray}

Note that this frequency does not depend on $T_{2}$. Hence, the BIC
which appears at this frequency does not come from the Fano interference
between the two transition branches corresponding to $T_{1}$ and
$T_{2}$. As discussed in \cite{Tanaka}, in this BIC the electron is trapped in a delocalized state extended over the two atoms and the section of wire between them.

Next we consider the case of Eq. (\ref{eq: dz}). This case leads
to a new type of BIC, which is a main result of the present paper.
In contrast to the BIC in Eq. (\ref{eq:Om}), the value of $\Omega$
that satisfies Eqs. (\ref{eq: dz}) and (\ref{eq:z}) must meet the
condition 
\begin{equation}
\Omega=E_{u}-E_{l}-\frac{BgT_{1}}{T_{2}}+(n+1)\frac{T_{1}T_{2}}{Bg}\label{eq:Om2}
\end{equation}
It should be noted that the frequency $\Omega$ depends on $g$ and
$T_{2},$ in contrast to the case Eq.(\ref{eq:Om}). 

Hence, we call this BIC  the dynamic BIC as mentioned in the introduction.
Note that in the limit $T_{2}\rightarrow0,$ the dynamic BIC disappears
for $T_{1}\ne0.$ Hence, the BIC is a result of existence of two transition
branches associated with $T_{1}$ and $T_{2}.$ In other words, the
BIC is a result of Fano interference.

It should be emphasized that all BICs obtained in our system exist
for any value of $E_{u}$ for a suitable value of $\Omega$ . This
is in contrast to the system without radiation field discussed in
\cite{Tanaka}, where the BIC occur only for special values given
by 
\begin{equation}
E_{u}=-B\cos\left(\frac{m\pi}{2x_{D}}\right)\label{eq:oldbic-1}
\end{equation}
In other words, the BICs in the system with decoupled lower $|D\rangle$
and $|A\rangle$ states occurs only for a special kind of\emph{ intra-atoms
}with the discrete state energies given by Eq.(\ref{eq:oldbic-1}).
In contrast, for the present system which $T_{1}\neq0$ and $T_{2}\neq0,$
the BICs in the system  may exist for any intra-atomic levels by
tuning the value of $\Omega$. In this sense, it is experimentally
more feasible to achieve the BIC in our system than the system we
have discussed in \cite{Tanaka}.

\section{{\normalsize BIC and General Solution of the Eigenvalue
EQ. (\ref{eq:disp})}}

\begin{figure}
\includegraphics[width=8cm]{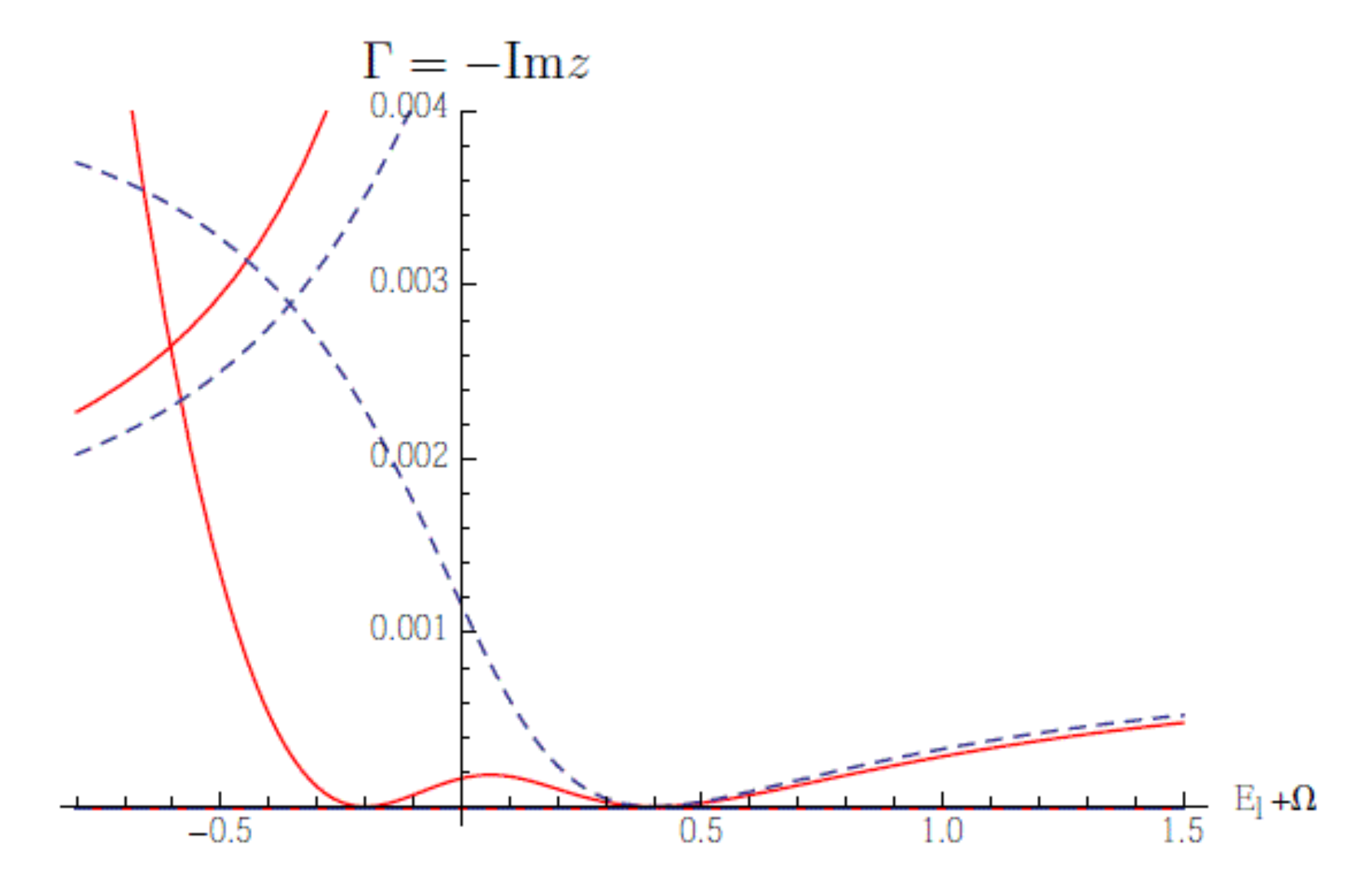}
\caption{\label{fig:n=00003D2g=00003D.2}Absolute value of the imaginary part
of eigenvalues of the Hamiltonian ($p$-sector) as a function of $\Omega+E_{l}$.
The parameters are $T_{1}=0.2,$ $g=0.2,$ $E_{u}=0.1$ and $x_{D}=2.$
The solid line corresponds to $T_{2}=0.2$, while the dashed line corresponds
to $T_{2}=0$. The curves on the upper-left corner correspond to another
solution of Eq.(\ref{eq:disp}). For $T_{2}=0.2$ (solid
line) there is a static BIC at $\Omega+E_{l}=0.4$ which is independent of the strength of the interaction $g$.  In addition there is a dynamic BIC at $\Omega+E_{l}=-0.2$ that is due to the interactin with the Fano interference.  The $\Omega+E_{l}$ values for which BIC occurs
in the plot are consistent with Eqs. (\ref{eq:Om}) and (\ref{eq:Om2}),
respectively. When $T_{2}=0$ (dashed line) the Fano interference is
suppressed, so only the first BIC occurs.}
\end{figure}

\begin{figure}
\includegraphics[width=8cm]{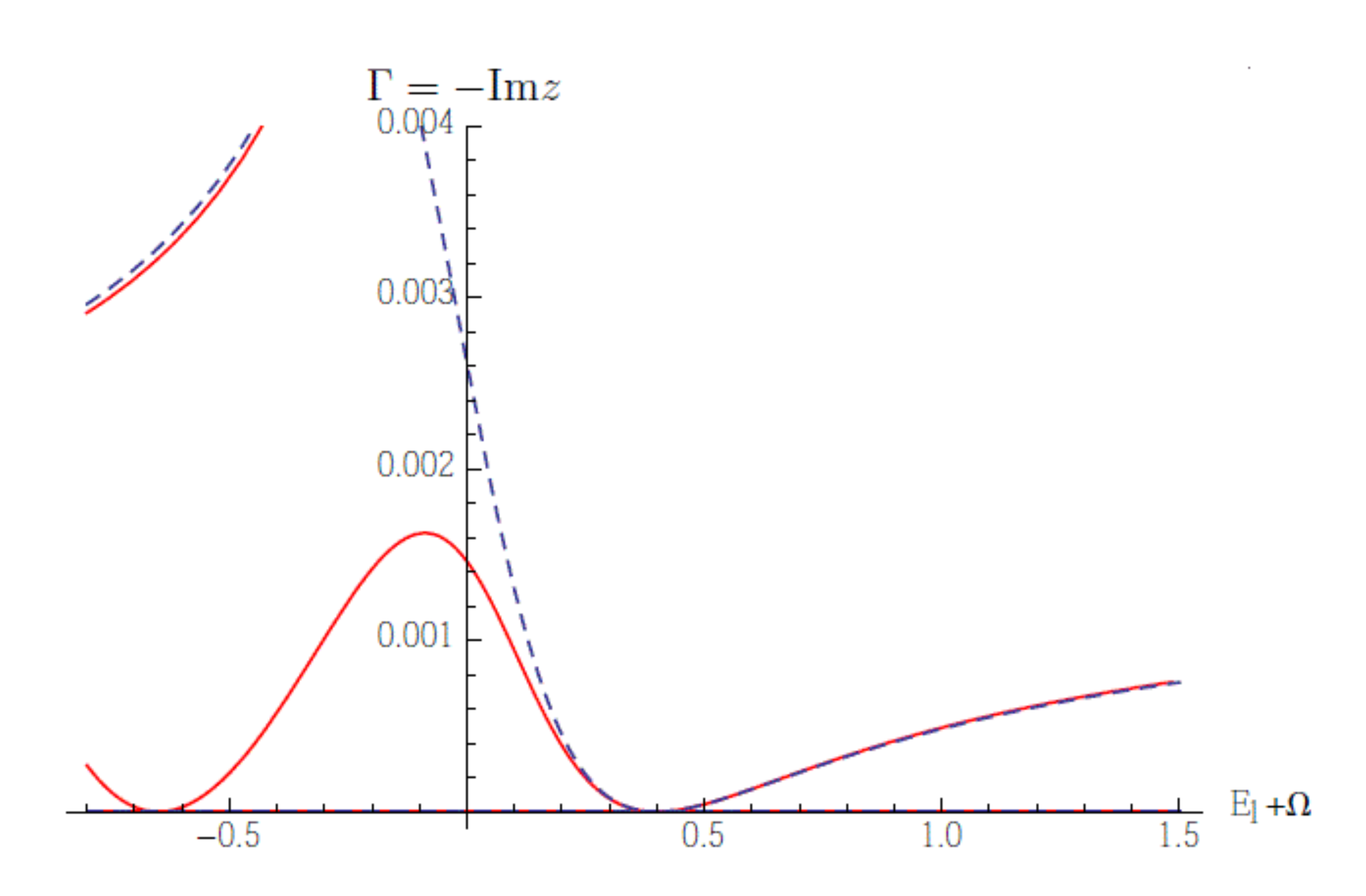}
\caption{\label{fig:n=00003D2g=00003D.4}Absolute value of the imaginary part
of eigenvalues of the Hamiltonian ($p-$sector) as a function of $\Omega+E_{l}$.
The parameters are the same as in figure \ref{fig:n=00003D2g=00003D.2}
except for $g=0.4.$ The solid line corresponds to $T_{2}=0.2$, 
while the dashed line corresponds to $T_{2}=0$.  The static BIC that occurs due to the vanishing of the self-energy still occurs at $\Omega+E_{l}=0.4$, while the dynamic  BIC is shifted to $\Omega+E_{l}=-0.659$ due to a change of $g$.}
\end{figure}

\begin{figure}
\includegraphics[width=8cm]{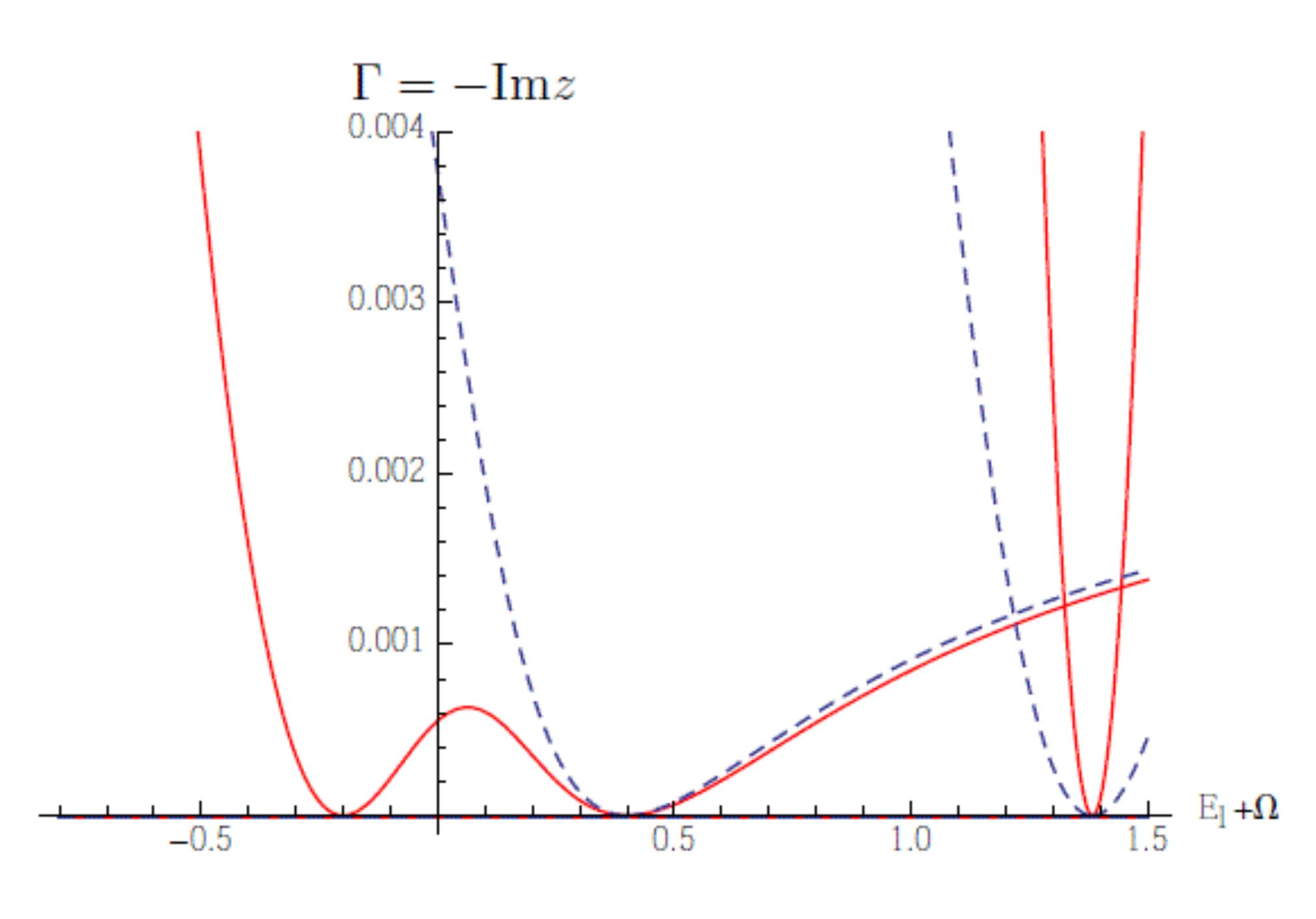}
\caption{\label{fig:n=00003D4g=00003D.2}Absolute value of the imaginary part
of eigenvalues of the Hamiltonian ($p$-sector) as a function of $\Omega+E_{l}$.
The parameters are the same as in figure \ref{fig:n=00003D2g=00003D.2}
except for $x_{D}=4.$ The solid line corresponds to $T_{2}=0.2$,
 while the dashed line corresponds to $T_{2}=0$. There are two BICs
at $\Omega+E_{l}=1.38$, and $\Omega+E_{l}=0.4,$ where the self-energy
vanishes. There is another BIC at $\Omega+E_{l}=-0.2$, for which
the self-energy does not vanish. }
\end{figure}

\begin{figure}
\includegraphics[width=8cm]{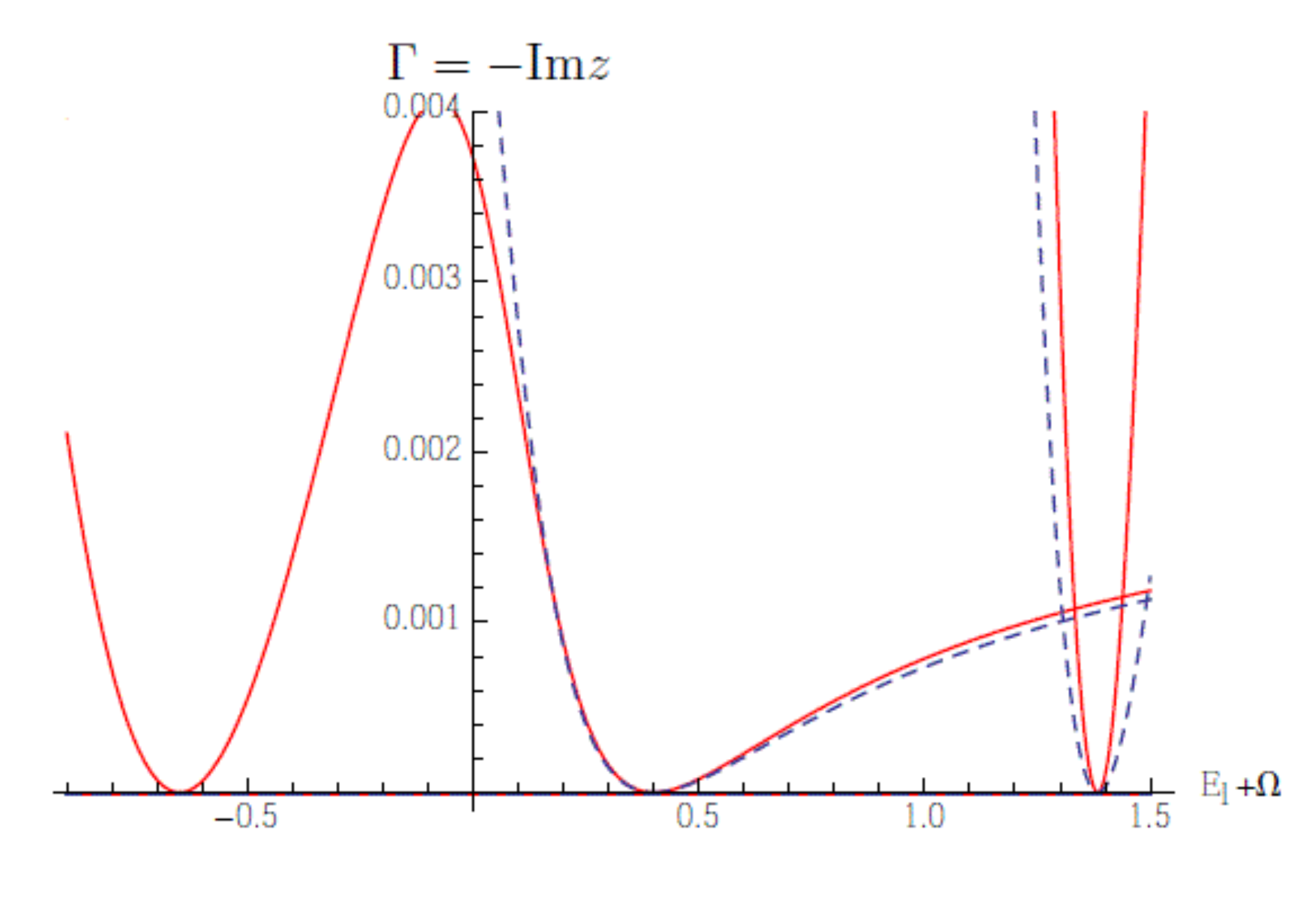}
\caption{\label{fig:n=00003D4g=00003D.4}Absolute value of the imaginary part
of eigenvalues of the Hamiltonian ($p$-sector) as a function of $\Omega+E_{l}$.
The parameters are the same as in figure \ref{fig:n=00003D4g=00003D.2}
except for $g=0.4.$ The solid line corresponds to $T_{2}=0.2$, 
while the dashed line corresponds to $T_{2}=0$. There are two static BICs at $\Omega+E_{l}=1.38$, and $0.4$,  while the dynamic BIC is shifted to $\Omega+E_{l}=-0.659$ due to a change of $g$. }
\end{figure}

In this section we will present numerical results showing  the general
solution of Eq. (\ref{eq:disp}) as function of $\Omega+E_{l}$ and
compare them to the analytic solutions of the BIC we obtained in the
previous section. For illustration we will consider the simplest case with $n = 0$  where Eq. (\ref{eq:Om})  reduces to a linear equation for  $\Omega$.  For  $n \not= 0$, there appear more static BICs than the simplest case  with $n = 0$.  However,  in order to demonstrate the essential difference between dynamic BIC and  static BIC, it is enough to show the simplest case. The numerical results were obtained through a numerical solution of Eq. (\ref{eq:disp}).  In FIGS. \ref{fig:n=00003D2g=00003D.2}-\ref{fig:n=00003D4g=00003D.4},
we plot the imaginary part of the solution, $\Gamma\equiv-\text{Im\ensuremath{z}}$
as a function of $\Omega+E_{l}$ for the $p$-sector. The figures
for the $s$-sector are essentially the same as except the locations
of BICs are different. 

In FIGS. \ref{fig:n=00003D2g=00003D.2}-\ref{fig:n=00003D4g=00003D.4},
we plot the case $E_{l}=0.1$ and $T_{2}=0.2.$ In all these figures
the red solid line corresponds to the case $T_{2}=0.2$, and the blue
dashed line corresponds to the case  $T_{2}=0$. We consider both cases in order to identify the BIC due to Fano interference.

We show in FIG. \ref{fig:n=00003D2g=00003D.2} the case $x_{D}=2$
and $g=0.2$; in FIG. \ref{fig:n=00003D2g=00003D.4} we have
the same $x_{D}=2$ but $g=0.4$. As theoretically predicted, we have
two BICs, one from Eq.(\ref{eq:Om}) and the other from Eq.(\ref{eq:Om2})
with $\Gamma=0$.

The BIC at the positive value of $\Omega+E_{l}$ is the static BIC
that exists even in the case $T_{2}=0.$ As one can see, the location
of the BIC is at same point in FIG. \ref{fig:n=00003D2g=00003D.2}
and FIG. \ref{fig:n=00003D2g=00003D.4}, though the value of $g$
is different. The BIC at the negative value of $\Omega+E_{l}$ in
FIG. \ref{fig:n=00003D2g=00003D.2} and FIG. \ref{fig:n=00003D2g=00003D.4}
is the dynamic BIC that exist only for the case $T_{2}\ne0.$ The
location of this BIC depends on the value of $g$ (compare FIG. \ref{fig:n=00003D2g=00003D.2}
and FIG. \ref{fig:n=00003D2g=00003D.4}). 

We show in FIG. \ref{fig:n=00003D4g=00003D.2} the case $x_{D}=4$
and $g=0.2$;  in FIG. \ref{fig:n=00003D4g=00003D.4} we show the case
with the same $x_{D}=4$ but $g=0.4$. As predicted, we have different
BICs: two are from Eq.(\ref{eq:Om}), and the other from Eq. (\ref{eq:Om2})
with $\Gamma=0.$

All
the static BICs are located at predicted values of $\Omega+E_{l}$.
They exist also in the case $T_{2}=0.$ The location of the static
BICs in FIG. \ref{fig:n=00003D4g=00003D.2} appear at the same points
in FIG. \ref{fig:n=00003D4g=00003D.4} though the value of $g$ is
different. The dynamic BIC appears at the negative values of $\Omega+E_{l}$
in FIG. \ref{fig:n=00003D4g=00003D.2}. We have this dynamic BIC only
for $T_{2}\ne0$. The location of the BIC depends on the value $g$
as predicted by Eq. (\ref{eq:Om2}).

\section{{\normalsize Summary }}

In this paper we have shown  tunable bound-states in continuum (BIC)
in a 1D quantum wire with two impurities, induced by an intense monochromatic
radiation field. We found a new type of BIC in this system that we call ``dynamic
BIC,'' in addition to the other type of BIC that we call ``static
BIC.'' In contrast to the static BIC, the energy of the dynamic BIC depends
on the coupling constant $g$  between the discrete state of the electron
and the continuous state of the electron. Moreover, we have shown
that the dynamic BIC occurs because of the Fano interference among
the two transition channels of the electron induced by the radiation
field. 

Furthermore, we have shown that all BICs obtained in our system exist for any value of $E_{l}$ of the discrete
state for a suitable frequency $\Omega$ of the radiation field. This is
not the case for the ordinary BIC without the radiation
field. In this sense, it is experimentally more feasible to achieve
the BIC in our system. 

In order to justify experimentally our theoretical results, however,
we need the Fano profile  of the absorption spectrum
of the radiation field \cite{Tanaka}. To construct the Fano profile, we have to
construct the eigenstates with the complex eigenvalue of the Hamiltonian
for the resonance states (see e.g. \cite{Bohm}). We hope to
present this  elsewhere.

\acknowledgements
We thank L.E. Reichl, J. Keto, A. Bohm and S. Garmon for insightful
discussions. Y.B. thanks the Robert. A. Welch Foundation (Grand No.
F-1051) for partial support of this work.


\begin{thebibliography}{10}

\bibitem[1]{Von -W} J. von Neumann and E. Wigner, Phys. Z. \noun{${\bf 30}$},
465 (1929).

\bibitem[2]{Sudarshan} E.C.G. Sudarshan,E. Tirapegui (ed.), Reidel Pub.
Co. (1981), pp. 237-245.

\bibitem[3]{F.H.Stillinger}F. H. Stillinger and D. R. Herrick, Phys.
Rev. A ${\bf 11}$, 446 (1975).

\bibitem[4]{G.Ordonez}G. Ordonez and S. Kim, Phys. Rev. A ${\bf 70}$,
032702 (2004).

\bibitem[5]{Tanaka} S. Tanaka, S. Garmon, G. Ordonez, and T. Petrosky,
Phy. Rev. B ${\bf 76}$, 153308 (2007).

\bibitem[6]{Longhi} S. Longhi,  Eur. Phys. J. B ${\bf 57}$,  45 (2007).

\bibitem[7]{Sadreev} A. Sadreev, E. Bulgakov and I. Rotter Phys. Rev. B ${\bf 73}$, 235342 (2006).

\bibitem[8]{F.Cappasso}F. Capasso, C. Sirtori, J. Faist, D. L. Sivco,
S.-N. G. Chu, and A. Y. Cho, Nature, London ${\bf 358}$, 565 (1992).

\bibitem[9]{P. S. Deo}P. S. Deo and A. M. Jayannavar, Phys. Rev.
B ${\bf 50}$, 11629 (1994).

\bibitem[10]{Ordonez-Na}G. Ordonez, K. Na, and S. Kim, Phys. Rev.
A ${\bf 73}$, 022113 (2006).

\bibitem[11]{Linda}H. Lee and L. Reichl, Phys. Rev. B. ${\bf 77}$,
205318 (2008).

\bibitem[12]{Tanaka0-1}S. Tanaka, S. Garmon, and T. Petrosky, Phys
Rev. B ${\bf 73},$ 115340 (2006).

\bibitem[13]{EIT}J Mompart and R. Corbalan, J. Opt. B: Quantum Semiclass.
Opt. 2 R7\textendash{}R24 (2000) .

\bibitem[14]{Pradhan11}N. Pradhan and D. D. Sarma, J. Phys. Chem.
Lett. \textbf{${\bf 2}$,} 2818 (2011).

\bibitem[15]{Chin09} P. T. K. Chin, J. W. Stouwdam, and R. A. J.
Janssen, nano Lett. ${\bf 9,}$ 745 (2009).

\bibitem[16]{Watanabe87} S. Watanabe and H. Kamimura, J. Phys. Soc.
Jpn. \textbf{${\bf 56},$} 1078 (1987).

\bibitem[17]{CohenTannouji} C. Cohen-Tannouji, J. Dupont-Roc, and
G. Grynberg, \textit{Atom-Photon Interactions,} (Willey-InterScience,
1992).

\bibitem[18]{Fano}U. Fano, Nuovo Cim. ${\bf 12}$, 156 (1935).

\bibitem[19]{Fano-1-1}U. Fano, Phys. Rev. ${\bf 124}$, 1866 (1961).

\bibitem[20]{Petrosky}T. Petrosky, S. Tasaki, and I. Prigogine, Physica
A ${\bf 173}$, 175 (1991).


\bibitem[21]{Bohm} A. Bohm, Quantum Mechanics: Foundations and Applications,
(Springer, 1986).

\end{thebibliography}
\end{document}